\documentclass{article}
\usepackage{amssymb}

\input{tcilatex}

\begin{document}

\title{Cosmological Bounds on Spatial Variations of Physical Constants}
\author{John D. Barrow \\
DAMTP, Centre for Mathematical Sciences,\\
Cambridge University, Wilberforce Rd., \\
Cambridge CB3 0WA, UK}
\maketitle

\begin{abstract}
We derive strong bounds on any possible large-scale spatial variation in the
values of physical 'constants' whose space-time evolution is driven by a
scalar field. These limits are imposed by the isotropy of the microwave
background on large angular scales in theories which describe space and time
variations in the fine structure constant, $\alpha $, the electron-proton
mass ratio, $\mu $, and the Newtonian gravitational constant, $G$.
Large-scale spatial fluctuations in the fine structure constant are bounded
by $\delta \alpha /\alpha \lesssim 2\times 10^{-9}$ and $\delta \alpha
/\alpha \lesssim 1.2\times 10^{-8}$ in the BSBM and VSL theories
respectively, fluctuations in the electron-proton mass ratio by $\delta \mu
/\mu \lesssim 9\times 10^{-5}$ in the BM theory and fluctuations in $G$ by $%
\delta G/G\lesssim 3.6\times 10^{-10}$ in Brans-Dicke theory. These derived
bounds are significantly stronger than any obtainable by direct observations
of astrophysical objects at the present time.

\end{abstract}

\section{Introduction}

The recent resurgence of interest in the possible slow variation of some
traditional 'constants' of nature has focussed almost exclusively upon their
time variation. This was led inspired to a considerable extent by the
capability of new astronomical instruments to measure spectral lines created
by the light from distant quasar to very high precision. The quasar data
analysed in refs. \cite{webb} using the new many-multiplet method consists
of three separate samples of Keck-Hires observations which combine to give a
data set of 128 objects at redshifts $0.5<z<3$. The many-multiplet technique
finds that their absorption spectra are consistent with a shift in the value
of the fine structure constant between these redshifts and the present of $%
\Delta \alpha /\alpha \equiv \lbrack \alpha (z)-\alpha ]/\alpha =-0.57\pm
0.10\times 10^{-5}$, where $\alpha \equiv $ $\alpha (0)$ is the present
value of the fine structure constant. Extensive analysis has yet to find a
selection effect that can explain the sense and magnitude of the
relativistic line-shifts underpinning these deductions. Further
observational studies have been published in refs. \cite{chand} using a
different but smaller data set of 23 absorption systems in front of 23
VLT-UVES quasars at $0.4\leq z\leq 2.3$ and have been analysed using an
approximate form of the many-multiplet analysis techniques introduced in
refs. \cite{webb}. They obtained $\Delta \alpha /\alpha \equiv -0.06\pm
0.06\times 10^{-5}$; a figure that disagrees with the results of refs. \cite%
{webb}. However, reanalysis is needed in order to understand the accuracy
being claimed. Other observational studies of lower sensitivity have also
been made using OIII emission lines of galaxies and quasars. The analysis of
data sets of 42 and 165 quasars from the SDSS gave the constraints $\Delta
\alpha /\alpha \equiv 0.51\pm 1.26\times 10^{-4}$ and $\Delta \alpha /\alpha
\equiv 1.2\pm 0.7\times 10^{-4}$ respectively for objects in the redshift
range $0.16\leq z\leq 0.8$ \cite{sdss}. Observations of a single quasar
absorption system at $z=1.15$ by Quast et al \cite{qu} gave $\Delta \alpha
/\alpha \equiv -0.1\pm 1.7\times 10^{-6}$, and observations of an absorption
system at $z=1.839$ by Levshakov et al \cite{lev} gave $\Delta \alpha
/\alpha \equiv 2.4\pm 3.8\times 10^{-6}$. A preliminary analysis of
constraints derived from the study of the OH microwave transition from a
quasar at $z=0.2467$, a method proposed by Darling \cite{darl}, has given $%
\Delta \alpha /\alpha \equiv 0.51\pm 1.26\times 10^{-4}$, \cite{oh}. A
comparison of redshifts measured using molecules and atomic hydrogen in two
cloud systems by Drinkwater et al \cite{drink} at $z=0.25$ and $z=0.68$ gave
a bound of $\Delta \alpha /\alpha <5\times 10^{-6}$ and an upper bound on
spatial variations of $\delta \alpha /\alpha <3\times 10^{-6}$ over 3 Gpc at
these redshifts; bounds on spatial variation of similar order arise from the
results of ref. \cite{webb} because of the wide distribution of the target
absorption systems on the sky.

New observational studies sensitive to small variations in the
electron-proton mass ratio, $\mu \equiv m_{e}/m_{p}$, at high redshift have
also been reported \cite{ubachs, petit, tz}, along with a new restriction on
possible time variation of the Newtonian gravitation constant, $G$, in the
solar system by the Cassini mission \cite{bert}. A range of other
astronomical and geophysical constraints which might limit possible changes
in $\alpha $ have also been re-evaluated \cite{uzan, olive, jdb}.

Theories which can handle space-time variations in $\alpha $ and $\mu $
self-consistently (as opposed to simply 'writing in' a time-varying constant
into the equations in which it is truly a constant) have only recently been
developed to complement the Brans-Dicke gravity theory \cite{bd} which has
been available to study the cosmological consequence of time-variations in $%
G $. The BSBM theory \cite{bek,bsbm} is a self-consistent extension of
general relativity which incorporates space-time variations in the $\alpha ,$
varying-speed-of-light (VSL) theories \cite{moff,vsl,vsl1,vsl2} provide
other ways of effecting variations in $\alpha $ and other gauge couplings,
and the new theory of Barrow and Magueijo \cite{bmmu} allows space-time
variations of $\mu $ to be studied. All of these theories model the
variation of a traditional constant by means of a scalar field which obeys a
conservation equation derived from the variation of an action. The time
variations of $\alpha ,\mu ,$ and $G$ that are permitted by these theories
have been investigated in varying degrees of detail. They differ in one
respect. The variations of the scalar fields carrying variations in $\alpha $
or $\mu $ do not have significant effects upon the expansion dynamics of the
universe: the latter remains well described by the usual general
relativistic FRW universe containing the appropriate matter source. However,
in the case of Brans-Dicke theory the changes in the associated scalar field
do affect the expansion dynamics of the universe and the FRW models are
changed into new solutions that approach those of general relativity only in
the limit that the space-time variation of $G$ tends to zero.

In this paper we will show how a simple treatment of these four
scalar-tensor theories for varying $\alpha ,\mu ,$ and $G$ allows us to
predict and bound the magnitude of the \textit{spatial} variations expected
in these constants in the universe on extragalactic scales by using the
observed temperature isotropy of microwave background radiation on large
angular scales. We note that, in the past, bounds on spatial variation have
been discussed by Tubbs and Wolfe \cite{tubbs}, and Pagel, \cite{pagel} who
addressed this question at a time when the large-scale uniformity of the
universe was a far greater mystery than it is in today's post-inflationary
era. These papers stressed that the values of combinations of physical
constants that were found to be the same to high precision when deduced from
the spectra of objects were so far apart on the sky that they could not have
been in causal contact during the history of the universe prior to the
emission of their light. Today, we expect a high degree of coherence within
the whole of the visible universe because it may have evolved from the
inflation of a single causally coherent domain. However, even if that were
the case, if constants like $\alpha ,\mu ,$ and $G$ are actually space-time
variables and possess small quantum statistical fluctuations at the time of
inflation then they may have a predictable (and even observable) spectrum of
inhomogeneous variations today. A particular example is given by the chaotic
inflationary universe in a Brans-Dicke theory of gravity \cite{linde,jb},
which gives rise to a spectrum of spatial fluctuations in the value of $G$
as well as in the density of matter.

In section 2 we shall describe four self-consistent theories of varying
constants and in section 3\ show how we can use the isotropy of the
microwave background in conjunction with the predictions of these theories
to derive bounds on the allowed spatial variations in these constants,
before summarising our results in section 4.

\section{Four representative theories}

We will consider four representative scalar theories that are of particular
interest given the current observational situation. It will be clear that
these theories have an analogous structure. In each case the conservation of
energy and momentum for the scalar field provides a wave equation of the form

\begin{equation}
\square \varphi =\lambda f(\varphi )L(\rho ,p),  \label{gen}
\end{equation}%
where $\varphi $ is a scalar field associated with the variation of some
'constant' $%
\mathbb{C}
$ via a relation $%
\mathbb{C}
=f(\varphi ),\lambda $ is a dimensionless measure of the strength of the
space-time variation of $%
\mathbb{C}
,$ $f(\varphi )$ is a function determined by the definition of $\varphi ,$
and $L(\rho ,p)$ is some linear combination of the density, $\rho $, and
pressure, $p$, of the matter that is coupled to the field $\varphi $ and $%
f(\varphi )\simeq 1$ for small $\varphi $. At a given cosmic time, for small
changes in $\varphi ,$ this equation describes small spatial variations in $%
\mathbb{C}
$ by a Poisson equation of the form

\begin{equation}
-\bigtriangledown ^{2}\left( \frac{\delta 
\mathbb{C}
}{%
\mathbb{C}
}\right) \simeq \lambda L(\rho ,p).  \label{master}
\end{equation}

\subsection{BSBM varying-$\protect\alpha $ theory}

A simple theory with time varying $\alpha $ was first formulated by
Bekenstein \cite{bek} as a generalisation of Maxwell's equations but
ignoring the consequences for the gravitational field equations. Recently,
this theory has been extended \cite{bsbm} to include the coupling to the
gravitational sector and some of its general cosmological consequences have
been analysed.

Variations in the fine structure 'constant' are driven explicitly by
variations in the electron charge, $e$, and the fine structure 'constant' is
given by

\[
\alpha \equiv e^{2\psi }, 
\]%
where the the scalar $\psi $ field obeys an equation of motion of the form (%
\ref{gen}): 
\begin{equation}
\square \psi =-\frac{2}{\omega _{1}}e^{-2\psi }\mathcal{\rho }_{em},
\label{boxpsi}
\end{equation}%
where $\omega _{1}$ is a dimensional constant which couples the kinetic
energy of the $\psi $ field to gravity and $\mathcal{\rho }_{em}$ is the
density of matter that carries electromagnetic charge. If we write $\zeta
\equiv \mathcal{\rho }_{em}/\rho $ where $\mathcal{\rho }_{m}$ is the total
matter density then

\[
\zeta =\frac{E^{2}-B^{2}}{E^{2}+B^{2}} 
\]%
and its sign ($-1\leq \zeta \leq 1$) depends on whether the dominant form of
(dark) matter is dominated by electrostatic ($E^{2}$) or magnetic ($B^{2}$)
energy (for further discussion see refs. \cite{bsbm, bkm, bs, sb, posp} .
For the scalar field, we have the propagation equation, 
\begin{equation}
\square \psi =-\frac{2}{\omega _{1}}e^{-2\psi }\zeta \rho _{m}
\label{psidot}
\end{equation}%
and for small variations in $\psi $ and $\alpha $ this is well approximated
by

\begin{equation}
\square \psi \simeq -\frac{2}{\omega _{1}}\zeta \rho _{m}  \label{bsbm}
\end{equation}

Some conclusions can be drawn from the study of the simple BSBM models with $%
\zeta <0$, \cite{bsbm}. These models give a good fit to the varying $\alpha $
implied by the quasar data of refs. \cite{webb}. There is just a single
parameter to fit to the data and this is given by the choice

\begin{equation}
\left\vert \frac{\zeta }{\omega _{1}}\right\vert =(2\pm 1)\times 10^{-4}
\label{om}
\end{equation}%
We shall use this as a conservative bound in what follows. Tighter
observational limits will only strengthen our conclusions.

\subsection{VSL theories}

In one 'covariant' version of these theories \cite{moff,vsl,vsl1} variations
in $\alpha $ (and all other gauge couplings $\alpha _{i}$) are driven by a
scalar field $\chi $ that drives explicit variations in the speed of light
and couples to all the matter fields in the lagrangian, not just the
electromagnetically-coupled matter, with $\alpha _{i}=\exp [Q\chi ],$ where $%
Q$ is a numerical constant. The structure of covariant VSL is analogous to
the BSBM theory and for small variations in the fine structure constant, $%
\exp [\chi ]\sim 1$, we have

\begin{equation}
\square \chi \simeq -\frac{Q}{\omega _{2}}\rho _{m}  \label{vsl}
\end{equation}%
where $\omega _{2}$ is a coupling constant. The observational data of Webb
et al \cite{webb} are fitted by

\[
\left\vert \frac{Q}{\omega _{2}}\right\vert =8\pm 4\times 10^{-4} 
\]%
which we use as the observational bound on the coupling. There is no
variation of $\alpha $ in the limit that $\frac{Q}{\omega _{2}}\rightarrow
0. $ Another edition \cite{vsl2} of a VSL theory has variation only in the
electromagnetic coupling, $\alpha ,$ and variations are driven only by the
pressure of matter:

\begin{equation}
\square \chi \simeq -4\pi G\omega p(\rho )  \label{vsl2}
\end{equation}%
for some new coupling constant $\omega $. We will just examine the covariant
VSL theory, eq. (\ref{vsl}) in what follows.

\subsection{BM varying-$\protect\mu $ theory}

The theory recently devised by Barrow and Magueijo \cite{bmmu} describes a
varying electron-proton mass ratio, $\mu $, via a changing electron mass
which is driven by a scalar field, $\phi $, defined by

\[
m_{e}=m_{0}e^{\phi } 
\]%
where $\phi $ obeys

\[
\square \psi =-\frac{m_{0}(n_{e}-n_{p})}{\omega _{3}}e^{\phi }.\mathcal{\ } 
\]%
Here, $n_{e}$ and $n_{p}$ are the electron and proton number densities, and $%
\omega _{3}$ is a dimensional coupling constant. In the case of small
variations ($e^{\phi }\sim 1$) this is well approximated by

\begin{equation}
\square \psi \simeq \square \mu \simeq -\frac{\rho _{e}}{\omega _{3}}
\label{mu}
\end{equation}%
and observational bounds on the time-variation of $\mu $ at high redshift 
\cite{tz} impose a weak bound of

\[
G\omega _{3}>0.2 
\]%
The $\omega _{3}\rightarrow \infty $ limit is that of constant $\mu $.

\subsection{Brans-Dicke gravity theory}

The Brans-Dicke theory \cite{bd} generalises Einstein's general theory of
relativity to incorporate a space-time variation in the Newtonian 'constant' 
$G$ by means of a Brans-Dicke scalar field $\Phi $ $\propto G^{-1}$ which
obeys a conservation equation of the form

\begin{equation}
\square \Phi =\frac{8\pi (\rho -3p)}{3+2\omega _{bd}},  \label{bd}
\end{equation}%
where $\rho $ and $p$ denote the total density and pressure of matter
respectively, $\omega _{bd}$ is the dimensionless Brans-Dicke parameter and
general relativity ($G=$ constant) is approached as $\omega _{bd}\rightarrow
\infty $. The current observational lower bounds on the allowed time
variation of $G$ give $\omega _{bd}$ $>500$ from a variety of local
gravitational tests (see \cite{uzan} for a review). But the strongest
constraint to date is derived from observations of the time delay of signals
from the Cassini spacecraft as it passes behind the Sun. These
considerations led Bertotti, Iess and Tortora \cite{bert}, after a
complicated data analysis process, to claim that 
\begin{equation}
\omega _{bd}>40000\text{ }(2\sigma ).  \label{bdbound}
\end{equation}

This theory differs from the three listed above in that small variations in $%
\Phi $ have a significant effect upon the expansion dynamics of the universe
because these variations control the strength of gravity. These variations
can be seen explicitly by writing down the Friedmann equation for the
expansion scale factor $a(t)$ in the case of zero spatial curvature:

\begin{equation}
\frac{\dot{a}^{2}}{a^{2}}=\frac{8\pi }{3\Phi }\rho -H\frac{\dot{\Phi}}{\Phi }%
+\frac{\omega _{bd}}{6}\frac{\dot{\Phi}^{2}}{\Phi ^{2}}\ .  \label{bdfrw}
\end{equation}%
For the case of dust ($p=0$) there are simple power-law solutions:

\begin{equation}
a(t)=t^{[2+2\omega _{bd}]/[4+3\omega _{bd}]}  \label{bds1}
\end{equation}

\begin{equation}
\Phi (t)=\Phi _{0}t^{2/[4+3\omega _{bd}]}  \label{bds2}
\end{equation}%
and we note that we recover the usual Einstein-de Sitter cosmology with $%
a=t^{2/3}$ as $\omega _{bd}\rightarrow \infty $. These solutions solve eq. (%
\ref{bd}) exactly in the $p=0$ case \cite{G}:

\QTP{Body Math}
\begin{equation}
\square \Phi =\frac{8\pi \rho }{3+2\omega _{bd}}.  \label{bd1}
\end{equation}%
\bigskip

\section{Large-scale inhomogeneity in $\protect\alpha ,\protect\mu ,$ and $G$%
}

The four theories that we have introduced have a characteristic structure in
which changes in $\alpha ,\mu ,$ and $G$ are driven by different parts of
the density of matter in the universe. This means that inhomogeneity in the
matter content of the universe is coupled to inhomogeneities in the
'constants' $\alpha ,\mu ,$ and $G.$ If we ignore the non-Machian mode that
is not driven by the matter fields (ie the complementary function arising
from the solution of $\square (scalar)=0$) because it falls off rapidly in
time and becomes negligible at late times in the universe, then we can
estimate the allowed large-scale spatial inhomogeneity of $\alpha ,\mu ,$
and $G$ in terms of observable quantities. The inhomogeneity in the
'constants' requires inhomogeneity in the driving matter perturbations and
their associated gravitational potential fluctuations. The magnitude of the
latter is observationally constrained by the temperature isotropy of the
microwave background on large angular scales and leads to an upper bound on
the possible inhomogeneity in the values of 'constants'. We shall ignore the
acceleration of the universe which began recently at $z\simeq 0.3.$ Its
inclusion leads to a very small change in the final results, no larger than
the uncertainty in other parameters.

\subsection{BSBM inhomogeneity in $\protect\alpha $}

On a hypersurface of constant comoving proper time, $t$, eq. (\ref{bsbm})
gives

\[
\nabla ^{2}(\delta \psi )\simeq \frac{\zeta }{\omega _{1}}\delta \rho _{m} 
\]%
So, noting that $6\pi G\rho t^{2}\simeq 1,$ any inhomogeneity in $\alpha $
over a scale $L$ is linked to inhomogeneity in the density of
electromagnetically-coupled matter, $\delta \rho _{em}$, and/or
inhomogeneity in the electromagnetic quality of the dark matter, $\delta
\zeta _{em}$, by \footnote{%
Note that $\triangledown ^{2}\psi \rightarrow \frac{10}{3L^{2}}\left( \psi
(x)-\bar{\psi}(x)\right) $ where $\bar{\psi}$ is the average value of $\psi $
in a spherical region of radius $L$ as $L\rightarrow 0$ and is defined by $%
\bar{\psi}(\vec{x})\equiv \frac{3}{4\pi R^{3}}\dint_{V}\psi (\vec{x}+\vec{r}%
)d^{3}\bar{\NEG{r}}$.}

\[
\delta \psi \simeq \frac{\delta \alpha }{\alpha }\simeq 0.3\frac{\zeta }{%
\omega _{1}}\left( \frac{L}{t}\right) ^{2}\left[ \frac{\delta \left( \zeta
\right) }{\zeta }+\frac{\delta \rho _{em}}{\rho _{em}}\right] . 
\]%
However, we expect that the inhomogeneity in the electromagnetically coupled
matter can be written in terms of the inhomogeneity of the total matter
density on scale $L$ in terms of some biasing parameter $\beta $ which will
not depart too greatly from being $O(1),$ so we put

\[
\frac{\delta \rho _{em}}{\rho _{em}}\simeq \beta \frac{\delta \rho _{m}}{%
\rho _{m}}. 
\]%
Hence, we have

\[
\frac{\delta \alpha }{\alpha }\simeq 0.3\frac{\zeta }{\omega _{1}}\left[ 
\frac{\delta \left( \zeta \right) }{\zeta }\left( \frac{L}{t}\right)
^{2}+\beta \frac{\delta \rho _{m}}{\rho _{m}}\left( \frac{L}{t}\right) ^{2}%
\right] , 
\]%
but we note that the gravitational potential perturbations ($\delta \Phi
_{N}/\Phi _{N}$) in the space-time metric created by $\delta \rho _{m}$ are
given by Poisson's equation ($\triangledown ^{2}\Phi _{N}=4\pi G\rho _{m}$)
as

\[
\frac{\delta \Phi _{N}}{\Phi _{N}}\simeq 0.3\frac{\delta \rho _{m}}{\rho _{m}%
}\left( \frac{L}{t}\right) ^{2} 
\]%
and these fluctuations produce temperature fluctuations in the microwave
background radiation, $\Delta T/T\simeq \delta \Phi _{N}/3\Phi _{N}\simeq
2\times 10^{-5}$, on the corresponding angular scales, \cite{sw} so that

\[
\frac{\delta \Phi _{N}}{\Phi _{N}}\simeq 0.3\frac{\delta \rho _{m}}{\rho _{m}%
}\left( \frac{L}{t}\right) ^{2}\simeq 3\frac{\Delta T}{T}\simeq 6\times
10^{-5} 
\]%
on large angular scales ($\theta >10^{0}$), \cite{cobe}. We will assume, in
accord with observations, that the gravitational potential fluctuations are
scale-independent to a good approximation \footnote{%
Note that the smallness of $\delta \Phi _{N}/\Phi _{N}$ is the justification
for the so called 'cosmological principle' and the use of the Friedmann
metric. The smallness of $\delta \rho _{m}/\rho _{m}$ is unnecessary \cite%
{cp}.}.

Hence, if we take the best fit to $\left\vert \frac{\zeta }{\omega _{1}}%
\right\vert \simeq 2\times 10^{-4}$ from eq. (\ref{om}) the observations of
ref. \cite{webb} and we assume that the spatial variations in the
electromagnetic composition of the matter in the universe are approximately
proportional to the variations in the matter density, with

\[
\frac{\delta \left( \zeta \right) }{\zeta }\lesssim \eta \frac{\delta \rho
_{em}}{\rho _{em}}, 
\]%
where $\eta \sim O(1),$ then the spatial fluctuations in $\alpha $ and the
required microwave background temperature anisotropies in these theories are
simply related by

\begin{equation}
\frac{\delta \alpha }{\alpha }\simeq 0.9\frac{\zeta }{\omega _{1}}\beta
(1+\eta )\frac{\Delta T}{T}\simeq 2\times 10^{-9}\beta (1+\eta ).
\label{bsbmlim}
\end{equation}

\bigskip Hence, for the two cases of small and large spatial variations in $%
\zeta ,$ respectively, we have

\begin{eqnarray*}
\frac{\delta \alpha }{\alpha } &\lesssim &\ 2\times 10^{-9}\beta \text{ if }%
\eta <<1, \\
\frac{\delta \alpha }{\alpha } &\lesssim &\ 2\times 10^{-9}\beta \eta \text{
if }\eta >1.
\end{eqnarray*}

\bigskip These bounds on large-scale spatial variations of the fine
structure 'constant' are extremely strong and we have obtained them by
assuming there is a bound on the level of time-variation consistent with the
observations of ref. \cite{webb}. For comparison, the sensitivity of direct
searches for \ variations in $\alpha $ which compare observations of
different quasar absorption spectra around the sky is only about $\delta
\alpha /\alpha \lesssim O(10^{-6})$. Recall that, in contrast, the time
variation of the fine structure constant is far more strongly constrained by
the quasar absorption-system data than by the microwave background effects
on small scales \cite{rocha}.

\subsubsection{VSL inhomogeneity in $\protect\alpha $}

A similar argument to that used for the BSBM case can be applied to the
covariant VSL theory and leads to the result that the allowed spatial
variation in $\alpha $ is again bounded by the temperature anisotropy in the
microwave background by

\[
\frac{\delta \alpha }{\alpha }\simeq 0.3\frac{Q}{\omega _{2}}\left[ \frac{%
\delta \rho _{m}}{\rho _{m}}\left( \frac{L}{t}\right) ^{2}\right] \simeq 0.9%
\frac{Q}{\omega _{2}}\frac{\Delta T}{T}\lesssim 1.2\times 10^{-8}. 
\]%
A similar bound could be deduced for the allowed spatial variations in all
the gauge couplings, $\delta \alpha _{i}/\alpha _{i}$ in this theory. Again,
this bound is far stronger than can be achieved by direct spectroscopic
studies of quasars and other astrophysical systems at low redshift.

\subsection{BM inhomogeneity in $\protect\mu $}

If we apply this argument to possible spatial variations in the
electron-proton mass ratio, $\mu $, described above then a similar argument
leads to an expression for the inhomogeneity in the electron-proton mass
ratio, $\mu ,$ of the form

\[
\frac{\delta \mu }{\mu }\simeq \frac{0.3}{G\omega _{3}}\left[ \frac{\delta
\rho _{e}}{\rho _{e}}\left( \frac{L}{t}\right) ^{2}\right] \simeq \frac{%
0.9\beta }{G\omega _{3}}\frac{\Delta T}{T}\lesssim \ 9\times 10^{-5\ }\beta 
\]%
where we have assumed that the inhomogeneity in the electron density is
approximately proportional to that in the matter distribution:

\[
\frac{\delta \rho _{e}}{\rho _{e}}=\beta \frac{\delta \rho _{m}}{\rho _{m}}. 
\]%
In this case the numerical bounds on the allowed variation are much weaker
than for $\delta \alpha /\alpha $. This is a reflection of the weak
constraints that exist on time-varying $\mu $ in these theories \cite{bmmu}
because of the different time-evolution of $\alpha $ and $\mu $ during the
radiation era.

\subsection{\protect\bigskip Brans-Dicke inhomogeneity in $G$}

The analysis of the level of inhomogeneity expected in a Brans-Dicke
universe is slightly different because time variations in $\Phi \sim G^{-1}$
determine the expansion dynamics. For the 'Machian' solutions (\ref{bds1})-(%
\ref{bds2}) that are the attractors at late times we have $8\pi \rho \simeq
\Phi H^{2}$ and so inhomogeneity in the Brans-Dicke field ($\delta \Phi \neq
0$) is linked to inhomogeneity in the gravitation 'constant' ($\delta G\neq
0 $) and in the total matter density ($\delta \rho \neq 0$) by

\[
\frac{\delta G}{G}\simeq \frac{\delta \Phi }{\Phi }\simeq \frac{10.8}{%
3+2\omega _{bd}}\frac{\delta \rho _{m}}{\rho _{m}}\left( \frac{L}{t}\right)
^{2}\frac{(\omega _{bd}+1)^{2}}{(4+3\omega _{bd})^{2}}\simeq \frac{10.8}{%
3+2\omega _{bd}}\ \frac{(\omega _{bd}+1)^{2}}{(4+3\omega _{bd})^{2}}\frac{%
\Delta T}{T}. 
\]%
Hence, for large values of $\omega _{bd}$, as observations of the
time-variation of $G$ require, this simplifies to

\[
\frac{\delta G}{G}\simeq \ \frac{6}{5\omega _{bd}}\ \times 10^{-5}\lesssim
3.6\ \times 10^{-10} 
\]%
if we use the Cassini bounds on $\omega _{bd}$. If we replace the Cassini
bound by an observational bound of $\omega _{bd}>1000$ from other
solar-system constraints this weakens the bound by a factor of 40. In either
case the bound on spatial variations is extremely strong. It results from a
combination of the microwave background anisotropy limits on the density
perturbations which drive variations in $G$ and the intrinsic weakness of
the $\Phi $ coupling. Notice that, just as fluctuations in the microwave
background temperature are known far more accurately than the mean
temperature itself, so the spatial fluctuations in $G$ are limited to far
greater accuracy than the value of $G$ is known experimentally (see for
example \cite{cod}).

We summarise the principal results obtained in Table 1.

\bigskip

\begin{tabular}{|l||l|l|l|l|}
\hline
{\small Theory} & {\small Scalar field} & {\small Scalar coupling} & {\small %
Scalar equation} & {\small Bound on inhomogeneity} \\ \hline\hline
{\small BSBM} & ${\small \psi :\alpha \equiv e}^{2\psi }$ & $\left\vert 
\frac{\zeta }{\omega _{1}}\right\vert {\small =2}_{-1}^{+1}{\small \times 10}%
^{-4}$ & ${\small \square \psi =-}\frac{2\mathcal{\rho }_{em}e^{-2\psi }}{%
\omega _{1}}$ & $\frac{\delta \alpha }{\alpha }{\small \simeq }\frac{\zeta }{%
\omega _{1}}{\small \beta \eta }\frac{\Delta T}{T}{\small \lesssim 2\times 10%
}^{-9}$ \\ \hline
{\small VSL} & ${\small \chi :\alpha }_{i}{\small \equiv e}^{Q\chi }$ & $%
\left\vert \frac{Q}{\omega _{2}}\right\vert {\small =8}_{-4}^{+4}{\small %
\times 10}^{-4}$ & ${\small \square \chi =-}\frac{Q\rho _{m}e^{-2\chi }}{%
\omega _{2}}$ & $\frac{\delta \alpha }{\alpha }{\small \simeq }\frac{Q}{%
\omega _{2}}\frac{\Delta T}{T}{\small \lesssim 1.2\times 10}^{-8}$ \\ \hline
{\small BM} & ${\small \phi :\mu \propto e}^{\phi }$ & $\left\vert G\omega
_{3}\right\vert {\small >0.2}$ & ${\small \square \phi =\ -}\frac{\rho
_{e}e^{\phi }}{\omega _{3}}$ & $\frac{\delta \mu }{\mu }{\small \simeq }%
\frac{\beta }{G\omega _{3}}\frac{\Delta T}{T}{\small \lesssim 9\times 10}%
^{-5}$ \\ \hline
{\small BD} & ${\small \Phi :G\propto \Phi }^{-1}$ & ${\small \omega }_{bd}%
{\small >40000}$ & ${\small \square \Phi =}\frac{8\pi \rho }{3+2\omega _{bd}}
$ & $\frac{\delta G}{G}{\small \simeq }\frac{3}{5\omega _{bd}}\frac{\Delta T%
}{T}{\small \lesssim 3.6\times 10}^{-10}$ \\ \hline
\end{tabular}

\bigskip

Table 1: Summary of the inhomogeneity levels allowed by the
microwave-background temperature isotropy in the theories discussed in this
paper. In column 2 the link between the scalar field and the 'constant' is
defined; in column 3 observational bounds on the dimensionless scalar
coupling constant appearing in the theory and discussed in the text are
given; the propagation equation for the scalar field is given in column 4
and in column 5 the bounds on any spatial variation in the associated
constant imposed by the bound on the coupling and the isotropy of the
microwave background ($\Delta T/T\leq 2\times 10^{-5}$) are summarised.

\section{\protect\bigskip Discussion}

We have shown that the characteristic structure of scalar theories for the
space-time variation of supposed constants of Nature enables us to use the
observed bounds on the couplings in the theories obtained from observational
searches for time variations in the associated 'constants' and the microwave
background isotropy to obtain very strong bounds on any spatial fluctuations
in these constants over large astronomical scales. Since the scalar fields
which carry the space-time variations of constants are driven by all (or
part) of the matter content of the universe there will always be density
inhomogeneities present which drive inhomogeneities in the constants.
However, the gravitational potential fluctuations associated with these
large-scale density perturbations show up in the microwave background
temperature anisotropy on large angular scales. Their amplitudes are
therefore bounded above and this bound in combination with limits on the
strength of the coupling of the scalar field leads to a series of very
strong bounds on possible spatial fluctuations in the associated constants.
We have calculated these bounds for four representative theories which are
of current interest. The basic argument is of wider application to other
dilaton theories and with small modifications the bounds can be modified to
include the small changes that arise because of a non-constant curvature
spectrum of density perturbations. They will be strengthened if the
underlying couplings in these theories are strengthened by future or present
observational studies.

Detailed consideration of the behaviour of fluctuations on smaller scales
would lead to a different range of constraints. We note that the limits
presented here apply to large scales and do not apply to the possible
variations in the values of 'constants' that can arise because of the
non-linear evolution of cosmic overdensities into clusters, galaxies and
planetary systems. These are not bounded by the isotropy of the microwave
background and still permit significant spatial variations to arise on small
scales in the local universe \cite{bmot}.

\textbf{Acknowledgements }I would like to thank T. Clifton, J. Magueijo, D.
Mota, M. Murphy, and J.K. Webb for helpful discussions.

\end{document}